\documentclass[twocolumn, prl, superscriptaddress]{revtex4-1}
\usepackage{amssymb}
\usepackage{amsmath}
\usepackage{epsfig}
\usepackage{color}
\usepackage{graphics, graphicx}
\usepackage{bbold}
\usepackage{psfrag}
\usepackage{mathcomp}
\usepackage{subfigure}
\usepackage{verbatim}
\usepackage{color}
\usepackage[colorlinks,citecolor=blue]{hyperref}

\begin{document}
\title{Generalized Dynamical Duality of Quantum Particles in One Dimension}
\author{Yu Chen}
\affiliation{Beijing National Laboratory for Condensed Matter Physics, Institute of Physics, Chinese Academy of Sciences, Beijing 100190, China}
\affiliation{School of Physical Sciences, University of Chinese Academy of Sciences, Beijing 100049, China}
\author{Xiaoling Cui}
\email{xlcui@iphy.ac.cn}
\affiliation{Beijing National Laboratory for Condensed Matter Physics, Institute of Physics, Chinese Academy of Sciences, Beijing 100190, China}
\date{\today}

\begin{abstract}
We prove a generalized dynamical duality for identical particles in one dimension (1D). Namely, 1D  systems with arbitrary statistics --- including bosons, fermions and anyons --- approach the same momentum distribution after long-time expansion from a trap, provided they share the same scattering length for short-range interactions. This momentum distribution is uniquely given by the rapidities, or quasi-momenta, of the initial trapped state. Our results can be readily tested in quasi-1D ultracold gases with tunable s- and p-wave interactions.
\end{abstract}

\maketitle

In the quantum world, different exchange statistics typically lead to distinct quantum phenomena. For example, in non-interacting systems, identical bosons tend to condense while identical fermions form a Fermi sea due to the Pauli exclusion principle. Even with interactions, bosonic and fermionic systems generally exhibit different quantum behaviors. In this context, identical particles in one dimension (1D) present a special case, as their energies and wavefunctions can be exactly mapped between different statistics under proper interaction strengths. This exact mapping (or duality) was first identified between hard-core bosons and non-interacting fermions~\cite{Girardeau1}, later extended to general couplings of bosons and fermions~\cite{Cheon} (also referred to as fermion-boson reciprocity~\cite{PP1977}), and recently generalized to anyons with fractional statistics~\cite{Cui, Blume1, Blume2}. The duality has taken the unique advantage of 1D geometry, where the wavefunction of quantum particles at a given spatial ordering does not depend on their statistics. Therefore, the generalized duality between different statistics only requires the same external potential and the same scattering length to characterize the short-range interaction~\cite{Cui}.

Despite equivalence in real and spectral space, 1D systems with different statistics can usually be distinguished by physical observables in momentum ($k$) space. This is because the $k$-space wavefunction depends on real-space configurations across different spatial orderings, and therefore exchange symmetry can strongly affect  $k$-space quantities. However, a recent cold atoms experiment observed an exceptional phenomenon~\cite{dyn_fer_expt}: the momentum distribution of hard-core bosons becomes identical to that of non-interacting fermions after long-time expansion from a trap. This is known as dynamical fermionization (DF), as predicted in both continuum~\cite{Gangardt, Campbell} and lattice systems~\cite{Rigol, Rigol2, Bolech}, and also extended to hard-core systems with fractional statistics~\cite{Campo, Piroli, Guo_expt}, spin degrees of freedom~\cite{Pu, Patu}, and finite temperatures~\cite{Patu2}. To explain this phenomenon, it has been shown for bosons that the long-time momentum distribution is related to the rapidity, or quasi-momentum, of the initial state~\cite{Campbell}, a conserved quantity in integrable systems. This explains why the hard-core bosons dynamically approach non-interacting fermions in $k$-space. Similar to DF, dynamical bosonization (DB) has also been revealed~\cite{Campo, Patu3, Minguzzi}, showing that identical fermions with infinitely strong $p$-wave attraction approach non-interacting bosons after long-time expansion.  Given the phenomena of DF and DB,  two important questions arise naturally: First, can such dynamical duality apply to general couplings of 1D system, and secondly, what are the roles of quantum statistics and coupling strengths in governing such a duality?   Motivated by the generalized duality in equilibrium~\cite{Cheon,Cui}, here we explore the possibility of generalized duality in dynamical systems. 

In this work, we exactly demonstrate a generalized dynamical duality between 1D systems with arbitrary statistics and general coupling strengths. We show that all 1D systems --- including bosons, fermions, and anyons --- approach the same momentum distribution after long-time expansion from a trap, given that they share the same scattering length for short-range interactions. The asymptotic momentum distribution is uniquely given by the quasi-momenta of the initial state before expansion. In this generalized duality manifold, DF~\cite{dyn_fer_expt,Gangardt, Campbell} and DB~\cite{Campo, Patu3, Minguzzi} constitute two special cases with zero and infinite scattering length, respectively. For a typically finite scattering length, we numerically confirm the dynamical duality between different statistics by exactly solving the dynamics of small clusters released from a harmonic trap. Our results can be readily detected in quasi-1D ultracold gases with tunable s- and p-wave interactions. The effect of a finite p-wave effective range in realistic cold atom experiments is also discussed.

We start from the  Hamiltonian of  identical particles in 1D with mass $m$ and coordinates $\{x_i\}$: ($\hbar=1$)
 \begin{eqnarray}
H&=&\sum_{i=1}^{N}\left(-\frac{1}{2m}\frac{\partial^2}{\partial x_i^2}+V_T(x_i)\right)+\sum_{i<j} U(x_j-x_i), \label{H}
 \end{eqnarray}
where $V_T$ is the external trapping potential, and $U$ is the short-range interaction that determines  the boundary condition at contact:
    \begin{equation}
 	\lim_{x\equiv x_j-x_i\rightarrow 0^+} \left( \frac{1}{l} + \partial_x\right) \Psi^{(\alpha)}(x_1,x_2,...x_N)=0. \label{Boundary_condition}
 \end{equation}
Here $\Psi^{(\alpha)}$ is the wavefunction of $N$ identical particles with statistics $\alpha$: for instance, bosons, fermions and anyons respectively correspond to $\alpha=0, \ \pm \pi$ and factional values between $0$ and $\pm \pi$.  In Eq.~(\ref{Boundary_condition}), the scattering length $l$   serves as the unique physical parameter to characterize short-range interaction strength for all $\alpha$ systems. Given a fixed $l$, all $\alpha$ systems share the same $\Psi^{(\alpha)}$ at a given spatial ordering (such as $x_1<x_2...<x_N$), while statistics $\alpha$ just determines the phase difference of $\Psi^{(\alpha)}$ between different ordering regimes. 
Explicitly, $\Psi^{(\alpha)}$ can be written as 
 \begin{equation}
\Psi^{(\alpha)}(x_{1}, ..., x_{N})=\sum_{Q} \theta(x_{Q1}<...<x_{QN})e^{i\frac{\alpha}{2}\Lambda(\vec{x}_Q)} \psi(\vec{x}_Q),  \label{general_wf}
 \end{equation}  
where $\vec{x}_Q=(x_{Q1},...,x_{QN})$; $\psi(\vec{x}_Q)$ is the wave function in spatial regime $x_{Q1}<...<x_{QN}$, as determined by the short-range boundary condition in Eq.~(\ref{Boundary_condition}). Setting the phase factor $\Lambda(\vec{x}_Q)=\sum_{j<k}^N\epsilon(x_j-x_k)$, with $\epsilon(x)=1$ for $x>0$ and $-1$ for $x<0$, we can see that $\Psi^{(\alpha)}$ satisfies the exchange symmetry required by statistics  $\alpha$:    
 \begin{equation}
	\Psi^{(\alpha)}(x_1,... x_j,...x_i,...x_N)=e^{i\alpha\omega}\Psi^{(\alpha)}(x_1,...x_i,...x_j,...x_N), \label{anyon_exchange}
\end{equation}
with $\omega=\sum_{k=i+1}^j \epsilon(x_k-x_i)-\sum_{k=i+1}^{j-1}\epsilon(x_k-x_j)$.  Since $\psi(\vec{x}_Q)$ in Eq.~(\ref{general_wf}) does not depend on $\alpha$, we can arrive at a generalized boson-anyon-fermion duality for equilibrium case~\cite{Cui}, i.e., all $\alpha$ systems with the same $l$ share the same energy spectrum and real-space density:
 \begin{eqnarray}
 H\Psi^{(\alpha)}_i(\vec{x})&=&E_i\Psi^{(\alpha)}_i(\vec{x}); \nonumber \\
 \rho_i(\vec{x})&=& |\Psi^{(\alpha)}_i(\vec{x})|^2. \label{duality}
  \end{eqnarray} 
Here $E_i$ and  $\rho_i$, both independent of $\alpha$, are respectively the energy and density distribution of the $i$-th eigenstate.  Note that the generalized duality in Eq.~(\ref{duality}) is robust against the choice of external potential ($V_T$) in Eq.~(\ref{H}). 

In principle, the duality cannot apply to momentum-space quantities, which involve particles moving across different spatial orderings and have to carry the information of $\alpha$. This makes the phenomenon of dynamical fermionization quite exceptional, which tells that the momentum distribution of hard-core bosons is identical to that of non-interacting fermions after long-time expansion from a trap~\cite{dyn_fer_expt,Gangardt, Campbell}. In the following, we will show that this dynamical phenomenon can in fact be generalized to general coupling strengths and to arbitrary statistics of 1D systems.

We consider the dynamics of identical particles after a sudden removal of  $V_T$  at time $t=0$. The scattering length is always taken to be negative ($l<0$), such that no bound state exists and the system expand freely after released from the trap. The free-space eigen-states can be exactly solved by the Bethe ansatz (BA) method~\cite{Guan_review}, as  labeled by a set of quasi-momenta $\{\vec{k}=(k_1,k_2,...k_N)\}$. These BA states follow the structure of Eq.~(\ref{general_wf}) and can be written as~\cite{Guan,Patu4,Chen2,Kundu} 
 \begin{equation}
\Phi^{(\alpha)}_{\vec{k}}(x_{1},..., x_{N})=\sum_{Q} \theta(x_{Q1}<...<x_{QN})e^{i\frac{\alpha}{2}\Lambda(\vec{x}_Q)}\phi_{\vec{k}}(\vec{x}_Q),  \label{BA}
 \end{equation}  
 with
 \begin{equation}
 \phi_{\vec{k}}(\vec{x}_Q)=\sum_{P} A(k_{P1},...,k_{PN}) e^{i\sum_{j=1}^N k_{Pj}x_{Qj}},
 \end{equation}
 where $A(k_{P1},...,k_{PN})=(-1)^Pe^{i\sum_{a<b}\tan^{-1}[(k_{Pa}-k_{Pb})l/2]}$ is determined by the short-range boundary condition Eq.~(\ref{Boundary_condition}). The eigen-energy of $\Phi^{(\alpha)}_{\vec{k}}$ is given by $E_{\vec{k}}=\sum_{j=1}^Nk_j^2/(2m)$. Further imposing a periodic boundary condition  (here $L$ is the system length)
 \begin{align}
\Phi^{(\alpha)}_{\vec{k}}(0, x_2,\ldots, x_{N}) =e^{-i\alpha(N-1)}\Phi^{(\alpha)}_{\vec{k}}(L, x_2,\ldots, x_{N}), \label{BA_boundary}
\end{align}
we arrive at the Bethe-ansatz equation for all $\alpha$ systems: 
\begin{eqnarray}
 	e^{ik_jL}=\prod_{j\neq l}\frac{k_j-k_l-2i/l}{k_j-k_l+2i/l}. \label{BA_equation}
 \end{eqnarray}
from which we can solve all quasi-momenta $\{\vec{k}\}$. Given the universal BA equation in Eq.~(\ref{BA_equation}), the solutions of $\{\vec{k}\}$ and $E_{\vec{k}}$ are independent of $\alpha$, a manifestation of boson-anyon-fermion duality in the exactly solvable framework~\cite{footnote}.

We now expand the initial state in terms of $\{\Phi^{(\alpha)}_{\vec{k}}\}$ and write its time evolution as
\begin{eqnarray}
	\Psi^{(\alpha)}(\vec{x},t)=\sum_{\vec{k}}c^{(\alpha)}(\vec{k})e^{iE_{\vec{k}}t}\Phi^{(\alpha)}_{\vec{k}}(\vec{x}),\label{expand}
\end{eqnarray}
with $\vec{x}\equiv (x_1,x_2,...,x_N)$ and $c^{(\alpha)}(\vec{k})$  the projection coefficient. Importantly, because the initial $\Psi^{(\alpha)}$ and the basis $\Phi^{(\alpha)}_{\vec{k}}$ follow the same phase structure in different ordering regimes (see Eqs.~(\ref{general_wf},\ref{BA})), $c^{(\alpha)}(\vec{k})$ does not depend on $\alpha$ and can be simplified as 
\begin{equation}
c(\vec{k})=\int d\vec{x} \sum_Q \theta(x_{Q1}<...<x_{QN}) \phi_{\vec{k}}^*(\vec{x}_{Q}) \psi(\vec{x}_{Q},t=0). \label{c_k}
\end{equation}
This shows $\alpha$-independent quasi-momentum distribution of initial state, which serves as an essential condition for generalized dynamical duality as  demonstrated below. 

Plugging Eq.~(\ref{BA}) into Eq.~(\ref{expand}), we have 
\begin{eqnarray}
&&	\Psi^{(\alpha)}(\vec{x},t)=\sum_{Q} \theta(x_{Q1}<...<x_{QN})e^{i\frac{\alpha}{2}\Lambda(\vec{x}_Q)}  \nonumber\\
&&\ \ \  \sum_{\vec{k}}c(\vec{k})\sum_{P} A(k_{P1},...k_{PN}) e^{i\sum_{j} [k_{Pj}x_{Qj}-k_{Pj}^2/(2m)]},  \label{expand_2}
\end{eqnarray}
At long time $t\rightarrow\infty$, $\Psi^{(\alpha)}$ can expand to large $\vec{x}$ that  grows linearly with $t$ and therefore the phase factor in Eq.~(\ref{expand_2})  oscillates rapidly with varying $\vec{k}$. This allows the application of stationary phase approximation (SPA)~\cite{spa}, which tells that the main contribution to  $k$-summation in Eq.~(\ref{expand_2}) comes from the stationary phase points 
\begin{equation}
	k_{Pj} = \frac{mx_{Qj}}{t}.\label{spa_1}
\end{equation}
This leads to the asymptotic wavefunction  
\begin{eqnarray}
\Psi^{(\alpha)}(\vec{x},t\rightarrow\infty)&&=\sum_{Q} \theta(x_{Q1}<...<x_{QN})e^{i\frac{\alpha}{2}\Lambda(\vec{x}_Q)}  \nonumber\\
&&c(\frac{m\vec{x}_{Q}}{t})A(\frac{m\vec{x}_{Q}}{t}) e^{i\sum_{j} mx_{Qj}^2/(2t)}.  \label{expand_3}
\end{eqnarray}
The Fourier transform of Eq.~(\ref{expand_3}) can then be obtained as  
\begin{eqnarray}
\Psi^{(\alpha)}(\vec{k},t\rightarrow\infty)&&=\sum_{Q} \theta(k_{Q1}<...<k_{QN})e^{i\frac{\alpha}{2}\Lambda(\vec{k}_Qt/m)}  \nonumber\\
&&\ c(\vec{k}_Q)A(\vec{k}_Q)e^{i\sum_jk_{Qj}^2t/(2m)},  \label{psi_k}
\end{eqnarray}
where $\vec{k}$ are the real momenta instead of quasi-ones. In obtaining this equation we have again applied SPA, which selects $x_{Qj}=k_{Qj}t/m$ at $t\rightarrow\infty$. Eqs.~(\ref{expand_3}, \ref{psi_k}) describe a physical picture  that after long-time expansion of 1D systems, the ordering of $\vec{x}$ in coordinate space and the ordering of $\vec{k}$ in momentum space have a one-to-one correspondence as  $\vec{x}=\vec{k}t/m$.   In this way, $\Psi^{(\alpha)}(\vec{k})$ follows the same structure as  $\Psi^{(\alpha)}(\vec{x})$, leading to the duality in  both real and momentum space.

Explicitly, from Eq.~(\ref{psi_k}) we can obtain the one-body momentum distribution
\begin{eqnarray}
n^{(\alpha)}(k,t\rightarrow\infty)&=& \int dk_2...dk_N |\Psi^{(\alpha)}(k,k_2...k_N;t\rightarrow\infty)|^2 \nonumber\\
&=& \int dk_2...dk_N |c(k,k_2,...k_N)|^2 \nonumber\\
&\equiv& n_{\rm q}(k,t=0). \label{dyn_duality}
\end{eqnarray}
To this end, we have demonstrated the generalized dynamical duality in 1D, namely, all $\alpha$ systems with the same $l$  approach the same momentum distribution  after a long-time expansion  from a trap, as given by the quasi-momentum distribution ($n_{\rm q}$) of initial state before expansion. 
This result substantially broadens the application scope of 1D duality, i.e., from equilibrium~\cite{Girardeau1, Cheon, Cui, Blume1, Blume2} to dynamical systems and from special statistics and coupling strengths~\cite{dyn_fer_expt,Gangardt, Campbell,Campo, Piroli, Patu3, Minguzzi} to general situations within a unified framework. 

\begin{widetext}

 \begin{figure*}[h]
	\includegraphics[width=18cm]{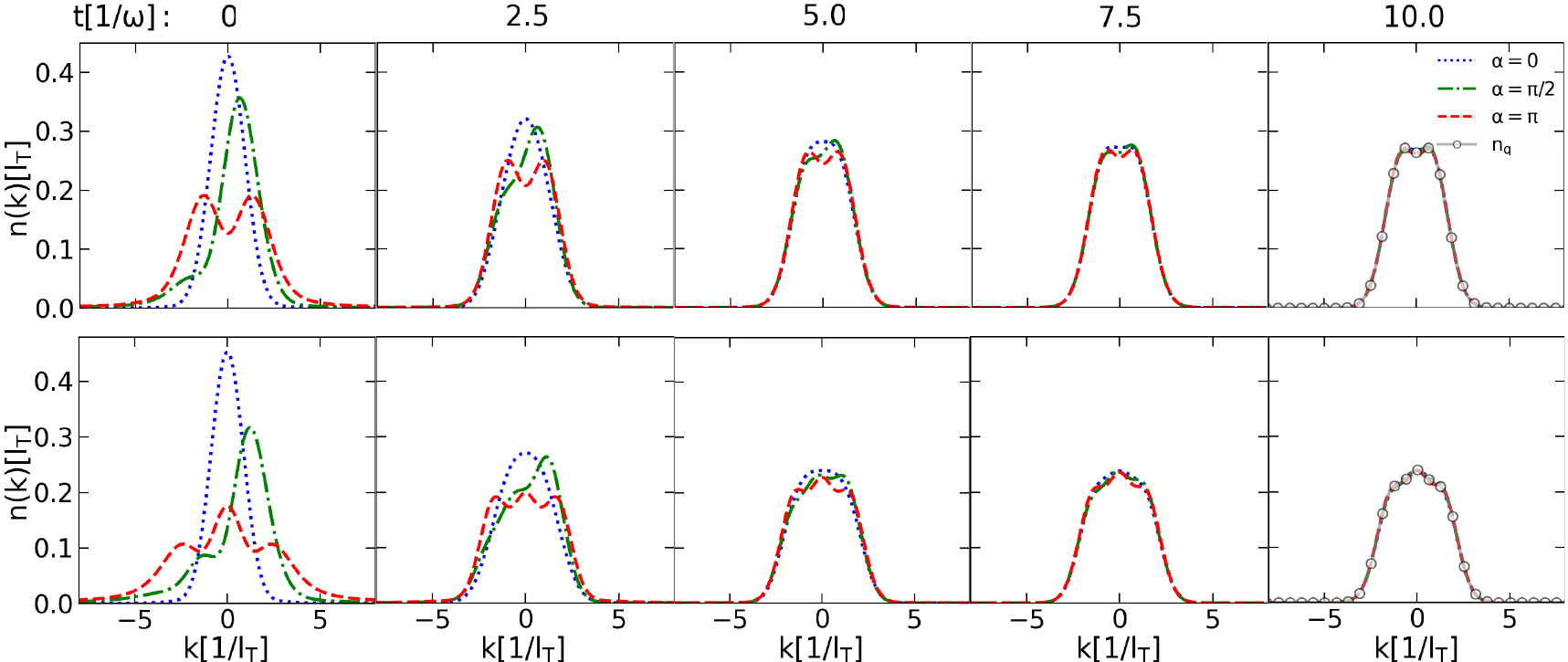}
	\caption{(Color online).  Momentum distributions of two (upper panel) and three (lower panel) identical particles at different expansion times after released from a harmonic trap. Here we take different statistics $\alpha=0$ (boson), $\pi/2$ (anyon) and $\pi$ (fermion), and the scattering length is fixed at $l=-l_{T}$, with $l_T=\sqrt{2/(m\omega)}$ the trap length and $\omega$ the harmonic frequency.  Gray lines with dots at the longest time ($t=10$) show quasi-momentum distributions of initial states. The units of  $k$, $n(k)$ and $t$ are respectively $1/l_T$, $l_T$ and $1/\omega$.} \label{fig_dyn}
\end{figure*}

 \end{widetext}

To confirm the generalized dynamical duality, we have performed exact calculations on expansion dynamics of small clusters after released from a harmonic trap $V_T(x)=m\omega^2x^2/2$. Specifically, we consider two and three identical particles with different statistics $\alpha=0$(boson), $\pi/2$(anyon) and $\pi$(fermion) at a typical scattering length $l=-l_T$ (here $l_T=\sqrt{2/(m\omega)}$ is the trap length). The initial state for each case is chosen as the ground state of trapped clusters, which can be exactly solved for s-wave bosons~\cite{Busch, DAmico} and p-wave fermions~\cite{Blume4,Cui3,Blume5}. More details of solving three-fermion problem using the renormalized p-wave interaction~\cite{Cui3} are presented in  ~\cite{supple}. 
As expected, these solutions respect the Bose-Fermi duality~\cite{Cheon}. Further, the initial state of anyons can be obtained by transforming the known bosonic ($\alpha=0$) or fermionic ($\alpha=\pi$) wavefunctions  to  fractional $\alpha$ based on Eq.~(\ref{general_wf}). Starting from each initial state, the dynamical evolution then follows Eq.~(\ref{expand}). In our numerics we have taken a large $L=80l_T$ and computed sufficiently many quasi-momentum states to expand Eq.~(\ref{expand})~\cite{supple}. The resulting momentum distributions $n(k)$ at various times during the dynamics are shown in Fig.~\ref{fig_dyn}.

From Fig.~\ref{fig_dyn}, we see that at initial time $t=0$,  the clusters display substantially different $n(k)$ for different $\alpha$. Specifically,  $n(k)$ for fermions ($\alpha=\pi$) is more extended in $k$-space than that for bosons ($\alpha=0$);  in contrast to bosons and fermions, $n(k)$ for anyons ($\alpha=\pi/2$) is strongly asymmetric between $k$ and $-k$~\cite{Cui, Blume1, Blume2,Chen2, Campo, Piroli, Guo_expt}. However, as time goes, these distinct $n(k)$ gradually converge and finally all merge into a single asymptotic curve at sufficiently long time. This curve is exactly the quasi-momentum distribution of  initial state, as shown by  gray lines with dots in the rightmost plots of Fig.~\ref{fig_dyn}. This verified the generalized dynamical duality in Eq.~(\ref{dyn_duality}). 
Our results show that the dynamical duality can be observed for small clusters after an expansion time of $t\sim10/\omega$,  corresponding to $\sim 16$ ms for typical $\omega=(2\pi)100$Hz in ultracold experiments. 

In ultracold gases, the s- and p-wave  interactions in quasi-1D can be conveniently tuned through confinement-induced resonances~\cite{Olshanii, Blume3, Pricoupenko, Tan}. However, a practical issue that may affect the experimental exploration is the presence of large effective range associated with a p-wave Fermi gas, which gives rise to  a finite p-wave range ($r_p$) in quasi-1D. Given a finite $r_p$, both the short-range boundary condition (Eq.~(\ref{Boundary_condition})) and the BA equation for p-wave fermions (Eq.~(\ref{BA_equation})) should be modified, by replacing $1/l$ by $1/l-r_p(mE)$ where $E$ is the pairwise collision energy  in center-of-mass frame~\cite{supple}. However, Eqs.~(\ref{expand}-\ref{dyn_duality}) remain  unchanged, leading to the robust conclusion that the momentum distribution of p-wave fermions after a long-time expansion still approaches the quasi-momentum distribution ($n_{\rm q}(k)$) of initial state before expansion. Therefore, we will just evaluate the effect of $r_p$ on $n_{\rm q}(k)$ of initially trapped system.

\begin{figure}[t]
	\includegraphics[width=8.5cm]{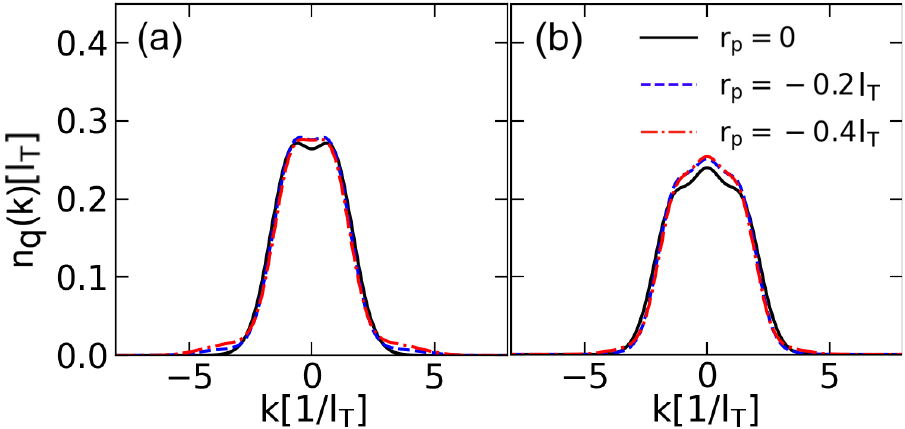}
	\caption{(Color online). Quasi-momentum distribution ($n_{\rm q}(k)$) of harmonically trapped two (a) and three (b) identical fermions at a fixed p-wave scattering length $l=-l_T$ and tunable effective range $r_p$. Here $l_T$ is the harmonic length, and the units of $k$, $n_{\rm q}$ and $r_p$ are respectively $1/l_T$, $l_T$ and $l_T$.} \label{fig_range}
\end{figure}

Fig.~\ref{fig_range} shows $n_{\rm q}(k)$ for two and three identical fermions at a fixed $l=-l_T$ and tunable $r_p$. One can see that a larger $|r_p|$ indeed leads to a larger deviation of $n_{\rm q}(k)$ from zero-range case. In quasi-1D, the reduced effective range follows $r_p=\frac{a_{\perp}^2 k_0}{12} - \frac{a_{\perp}}{4} \zeta(1/2, 1)$~\cite{Cui2}, where $k_0$ is the 3D p-wave range, $a_{\perp}$ is the transverse trap length and $\zeta(.,.)$ is the Hurwitz zeta function. For a realistic $^{40}$K Fermi gas with $k_0= -0.04a_0^{-1}$ ($a_0$ is the Bohr radius)~\cite{K40} and typical $a_{\perp}=70$nm, we have $r_p\sim -280$nm. Given the 1D trap length $l_T=10-20 a_{\perp}$, we have  $|r_p|/l_T\sim 0.2-0.4$. Within this range, Fig.~\ref{fig_range} shows that $n_{\rm q}(k)$ are only slightly modified from $r_p=0$ case, with main modifications in small-$k$ regime.  Therefore we expect the dynamical Bose-Fermi duality can be feasibly observed in realistic quasi-1D ultracold gases, even the Fermi gas is with a finite effective range.  
To explore the duality of anyons, one may first prepare a trapped anyonic gas following the s-p hybridization scheme~\cite{Cui} and then release it to measure $n(k)$. 

In summary, we have established a generalized dynamical duality for 1D quantum particles with arbitrary statistics. It tells that all 1D systems with the same scattering length, including bosons, fermions and anyons, approach the same momentum distribution  long after being released from a trap. This momentum distribution is uniquely given by the quasi-momentum distribution of initial state. These predictions can be practically detected  in quasi-1D ultracold gases with tunable interactions.  By extending the phenomena of DF~\cite{dyn_fer_expt,Gangardt, Campbell} and DB~\cite{Campo, Patu3, Minguzzi}  to arbitrary statistics and coupling strengths, our current study has significantly broadened the scope of  exact duality in 1D, and moreover, pointed to an intrinsically deep connection between 1D systems in a general context. 

In future, it will be interesting to examine the possibility of generalized dynamical duality in lattice systems, given the special case of DF established there~\cite{Rigol, Rigol2, Bolech}. %Inspired by our work, one may expect the lattice DF~\cite{Rigol, Rigol2, Bolech} can be equally generalized to arbitrary statistics and general couplings. 
In particular, the lattice version of anyons have recently been realized in cold atoms~\cite{anyon_expt}, opening up practical possibilities to explore the generalized  duality in related setup. Besides, our generalized duality may also be extended to spin mixtures and to finite temperatures, in view of the existence of DF therein~\cite{Pu, Patu, Patu2}.

Data that support the findings of this article are openly available~\cite{data}.

{\it Acknowledgements.} This work is supported by National Natural Science Foundation of China (12525412, 92476104, 12134015) and Quantum Science and Technology-National Science and Technology Major Project (2024ZD0300600).

\clearpage

\onecolumngrid
\vspace*{1cm}
\begin{center}
	{\large\bfseries Supplementary Materials}
\end{center}
% set prefix
\setcounter{figure}{0}
\setcounter{equation}{0}
\renewcommand{\figurename}{Fig.}
\renewcommand{\thefigure}{S\arabic{figure}}
\renewcommand{\theequation}{S\arabic{equation}}

In this supplementary material, we provide more details on the exact solutions of small clusters in trapped and continuum systems, as well as the effect of finite p-wave effective range.

\section*{I.\ \ \ Exact solutions of small clusters in trapped and continuum systems}

For small clusters confined in a 1D harmonic trap, previous studies have presented exact two-body solutions for identical bosons with s-wave interaction~\cite{Busch,DAmico} and for identical fermions with p-wave interaction~\cite{Blume4,Cui3, Blume5}. In particular, in Ref.~\cite{Cui3} we have established the renormalization equation for 1D p-wave coupling and utilized it to solve the problem of two identical fermions in a harmonic trap. In the following we will use the same method to solve three-fermion problem.   

For three identical fermions with coordinates $\{x_1,x_2, x_3\}$, we can separate out the center-of-mass (CoM) motion and define two relative coordinates as
\begin{equation}
	r=x_2-x_1,\ \ \ \ \rho=\frac{2}{\sqrt{3}}(x_3-\frac{x_1+x_2}{2}).
\end{equation}
Similarly, we have another two sets of relative coordinates:   $\{r_+,\rho_+\}$ and  $\{r_-,\rho_-\}$, by transforming $ \{r,\rho\}$ under particle exchanges   $x_2\leftrightarrow x_3$ and $x_1\leftrightarrow x_3$. In the CoM frame, the Hamiltonian can be written as $H(r,\rho)=H^{(0)}+U$, with
\begin{eqnarray}
	H^{(0)}&=&-\frac{1}{m}\left( \frac{\partial^2}{\partial r^2} +\frac{\partial^2}{\partial \rho^2} \right) + \frac{m}{4} \omega^2 (r^2+\rho^2); \\
	U&=& V(r)+V(r_+)+V(r_-).
\end{eqnarray}
Here the p-wave interaction potential is given by $V(r)=g \overleftarrow{\partial}_r \delta(r) \overrightarrow{\partial}_r$, where the bare coupling $g$ is related to the p-wave scattering length $l$ via the renormalization equation~\cite{Cui3}:
\begin{equation}
	\frac{1}{g}=\frac{m}{2l}-\frac{1}{L}\sum_k \frac{k^2}{2\epsilon_k},  \label{RG}
\end{equation}
with $\epsilon_k=k^2/(2m)$ and $L$ the length of the system.

The three-body wave function in CoM frame can be expanded as 
\begin{equation}
	\Psi(r,\rho)=\sum_{mn} c_{mn}\phi_{m}(r)\phi_{n}(\rho), \label{function_expansion}
\end{equation}
with single-particle eigen-state  
\begin{equation}
	\phi_{n}(x)=\frac{1}{\pi^{\frac{1}{4}}\sqrt{2^nn!l_T}}e^{-\frac{x^2}{2l_T^2}}H_n(x/l_T), \ \ \ \ (l_T=\sqrt{2/(m\omega)})
\end{equation}
and eigen-energy $\epsilon_n=(n+1/2)\omega$. 

Introducing an auxiliary function $f(r,\rho)\equiv U\Psi(r,\rho)$, and ensuring its anti-symmetry 
\begin{equation}
	f(r,\rho)=-f(r_+,\rho_+)=-f(r_-,\rho_-), \label{f_symmetry}
\end{equation} 
we can write $f$-function as  
\begin{equation}
	\begin{split}
		f(r,\rho)=g\sum_{mn}c_{mn}\phi_m'(0)\left(\overleftarrow{\partial}_r \delta(r)\phi_n(\rho)-\overleftarrow{\partial}_{r_+} \delta(r_+)\phi_n(\rho_+)-\overleftarrow{\partial}_{r_-} \delta(r_-)\phi_n(\rho_-)\right).
	\end{split} \label{f}
\end{equation}
Utilizing  the Lippmann-Schwinger equation 
\begin{equation}
	\Psi(r,\rho)=\int dr' d\rho' \langle r, \rho |G_0| r', \rho'\rangle f(r', \rho'), \label{LS}
\end{equation}
with $G_0=(E-H^{(0)})^{-1}$, and plugging Eqs.~(\ref{function_expansion},\ref{f}) into this equation, we obtain 
\begin{equation}
	\frac{1}{g}(E-\epsilon_m-\epsilon_n)c_{mn}=\sum_{ij}c_{ij}\phi_i'(0)(\phi_m'(0)\delta_{j,n}-C^+_{mn,j}-C^-_{mn,j}),  \label{fermi_eigen_equation1}
\end{equation}
where \begin{eqnarray*}
	C^{\pm}_{mn,j}=\int dx \left((\frac{1}{2}\phi'_m(\pm \sqrt{3}x/2)\phi_n(-x/2)\pm \frac{\sqrt{3}}{2}\phi_m(\pm\sqrt{3}x/2)\phi'_n(-x/2)\right)\phi_j(x).
\end{eqnarray*}
By defining $a_{n}=\sum_{m} c_{mn}\phi'_m(0)$, Eq.~(\ref{fermi_eigen_equation1}) can be further simplified as
\begin{equation}
	\begin{split}
		\left( \frac{1}{g}- \sum_{m}\frac{|\phi'_m(0)|^2}{E-\epsilon_m-\epsilon_n}\right) a_n=-\sum_ja_j\sum_{m}\frac{\phi'_m(0)(C^{+}_{mn,j}+C^{-}_{mn,j})}{E-\epsilon_m-\epsilon_n}.
	\end{split} \label{fermi_simplified_eigen_equation}
\end{equation}
From this equation one can obtain the eigen-energy $E$, and further the coefficients $\{c_{mn}\}$ can be obtained from Eq.~(\ref{fermi_eigen_equation1}). One can also prove from Eq.~(\ref{LS}) that the anti-symmetry of $\Psi$ under particle exchange can be automatically guaranteed by the anti-symmetry of $f$-function (see Eq.~(\ref{f})). 

Note that in the left side of Eq.~(\ref{fermi_simplified_eigen_equation}), both two terms in the bracket  have ultraviolet divergences at high energy, which can be exactly eliminated with each other and give rise to a physical result after subtraction. In fact, this divergence is related to the singularity (discontinuity) of p-wave wavefunction when two fermions get close to each other~\cite{Cui3}, and therefore is a universal fact regardless of the application of external trap. In our numerics, we have confirmed the convergence of our results by choosing different energy cutoffs in solving the matrix equation in Eq.~(\ref{fermi_simplified_eigen_equation}). The largest cutoff of $n$, which determines the  matrix size,  is taken to be $n_{\rm max} =80$, which allows the convergence of ground state energy up to the sixth digit (in unit of $\omega$).

For small clusters in continuum, a key issue is to obtain the quasi-momenta by solving the BA equations (Eq.~(9) in the main text). For two particles ($N=2$), setting $k_1=K/2-k$ and $k_2=K/2+k$ and taking the logarithm of both sides of BA equations, we obtain:
\begin{equation}
	\begin{aligned}
		KL&=2\pi N,\\
		kL+2\pi(n+\frac{1-N}{2})&=2\tan^{-1}(kl),
	\end{aligned}
	\label{ba_2}
\end{equation}
where $N$ and $n$ are integer quantum numbers corresponding to $K$ and $k$, respectively. Here $K$ is the CoM momentum and $k$ is the relative quasi-momentum.

For three particles ($N=3$), setting $k_1=K/3-k_r-k_\rho/\sqrt{3}$, $k_2=K/3+k_r-k_\rho/\sqrt{3}$ and $k_3=K/3+2k_\rho/\sqrt{3}$, and taking the logarithm of both sides of BA equations, we obtain:
\begin{equation}
	\begin{aligned}
		KL&=2\pi N,\\
		k_rL-n\pi&=2\tan^{-1}(k_rl)+\tan^{-1}(\frac{k_r+\sqrt{3}k_\rho}{2}l)+\tan^{-1}(\frac{k_r-\sqrt{3}k_\rho}{2}l),\\
		\frac{k_\rho L}{\sqrt{3}}-(m-N/3)\pi&=\tan^{-1}(\frac{k_r+\sqrt{3}k_\rho}{2}l)-\tan^{-1}(\frac{k_r-\sqrt{3}k_\rho}{2}l)
	\end{aligned}
	\label{ba_3}
\end{equation}
where $N$, $n$ and $m$ are integer quantum numbers defining $K$, $k_r$ and $k_\rho$ respectively. Here again $K$ is CoM momentum, and $k_r,\ k_{\rho}$ are two quasi-momenta corresponding to the motions of $r, \ \rho$ respectively.  In our numerics, we have imposed a cutoff of $[-60, 60]$ on the integer quantum numbers $N$, $n$, and $m$ to ensure the convergence of the results. After this, the BA basis $\Phi^{(\alpha)}_{\vec{k}}$ can be obtained using Eqs.~(6,7) in the main text. Note that $\Phi^{(\alpha)}_{\vec{k}}$ need to be normalized before utilized as a basis to expand $\Psi^{(\alpha)}$.

%For the calculation of anyons, we set their wave functions to be identical to those of s-wave bosons and p-wave fermions in the region where $x_1<x_2<...<x_N$, while the wave functions in other regions are directly obtained from exchange statistics.

\section*{II.\ \ \ Effect of a finite p-wave effective range}

Under a finite p-wave effective range $r_p$, both the short-range boundary condition and the BA equations (Eqs.~(2,9) in the main text) for p-wave fermions  should be modified, by replacing $1/l$ with $1/l-r_p(mE)$ where $E$ is the pairwise collision energy  in CoM frame. In Ref.~\cite{Cui3} we have discussed the effect of a finite $r_p$ to two-body solutions of trapped fermions. For three trapped fermions, one has to modify Eq.~(\ref{fermi_simplified_eigen_equation}) by replacing $1/g$  with 
\begin{equation}
	\frac{1}{g}\rightarrow \frac{m}{2} \left( \frac{1}{l}-r_p m (E-\epsilon_n) \right)-\frac{1}{L}\sum_k \frac{k^2}{2\epsilon_k}.
\end{equation}

For continuum system, the BA  equation (Eq.~(9) in the main text) should be modified as
\begin{eqnarray}
	e^{ik_jL}=\prod_{j\neq l}\frac{k_j-k_l-2i(1/l-r_p(k_j-k_l)^2/4)}{k_j-k_l+2i(1/l-r_p(k_j-k_l)^2/4)}. \label{BA equation_r}
\end{eqnarray}
Accordingly, for two fermions ($N=2$), Eq.~(\ref{ba_2}) becomes
\begin{equation}
	\begin{aligned}
		KL&=2\pi N,\\
		kL+2\pi(n+\frac{1-N}{2})&=2\tan^{-1}(\frac{kl}{1-k^2r_pl}),
	\end{aligned}
	\label{ba_2_r}
\end{equation}
and for three fermions ($N=3$), Eq.~(\ref{ba_3}) becomes
\begin{equation}
	\begin{aligned}
		KL&=2\pi N,\\
		k_rL-n\pi&=2\tan^{-1}(\frac{k_rl}{1-k_r^2r_pl})+\tan^{-1}(\frac{(k_r+\sqrt{3}k_\rho)l}{2-r_pl(k_r+\sqrt{3}k_\rho)^2/2})+\tan^{-1}(\frac{(k_r-\sqrt{3}k_\rho)l}{2-r_pl(k_r-\sqrt{3}k_\rho)^2/2})\\
		\frac{k_\rho L}{\sqrt{3}}-(m-N/3)\pi&=\tan^{-1}(\frac{(k_r+\sqrt{3}k_\rho)l}{2-r_pl(k_r+\sqrt{3}k_\rho)^2/2})-\tan^{-1}(\frac{(k_r-\sqrt{3}k_\rho)l}{2-r_pl(k_r-\sqrt{3}k_\rho)^2/2}).
	\end{aligned}
	\label{ba_3_r}
\end{equation}
Eqs.~(\ref{ba_2_r}, \ref{ba_3_r}) determine the modified quasi-momenta solutions in the presence of a finite $r_p$.


\begin{thebibliography}{99}
\bibitem{footnote} Note that the universality of Bethe-ansatz equation for all $\alpha$ does not depend on the choice of boundary condition. For instance, Ref.~\cite{Cui} adopted an open boundary condition and equally resulted in a universal Bethe-ansatz equation for all $\alpha$ systems.
\end{thebibliography}
\end{document}